# Evanescent-wave and open-air chiral sensing via signal-reversing cavity-enhanced polarimetry

Dimitris Sofikitis,[1,2*] Lykourgos Bougas,[1,2*] Georgios E. Katsoprinakis,[1,2] Alexandros K. Spiliotis,[1,2] Benoit Loppinet[1] and T. Peter Rakitzis[1,2]

[1]Institute of Electronic Structure and Laser, Foundation for Research and Technology-Hellas, 71110 Heraklion-Crete, Greece, [2]Department of Physics, University of Crete, 71003 Heraklion-Crete, Greece. *These authors contributed equally.

**Sensing chirality is of fundamental importance to many fields, including analytical and biological chemistry, pharmacology[1], and fundamental physics[2]. Applications range from drug design and synthesis, the determination of protein structure, to the measurement of parity violation of the weak force. Recent developments have extended optical chiral sensing using microwaves[3], fs pulses[4], superchiral light[5], and photoionization[6]. The most widely used methods are the traditional methods of circular dichroism and optical rotation. However, chiral signals are typically very weak, and their measurement is limited by larger time-dependent backgrounds (such as spurious birefringence) and by imperfect and slow subtraction procedures[7]. Here, we demonstrate a pulsed-laser bowtie-cavity-enhanced polarimeter with counter-propagating beams, which solves these background problems: the chiral signals are enhanced by the number of cavity passes (typically $>10^3$); the effects of linear birefringence are suppressed by a large induced intracavity Faraday rotation; and rapid signal reversals are effected by reversing the Faraday rotation and subtracting signals from the counter-propagating beams. These advantages allow measurements of absolute chiral signals in environments where background subtractions are not feasible: we measure optical rotation from chiral vapour in open air, and from chiral liquids in the evanescent wave produced by total internal reflection at a prism surface. Evanescent-wave optical rotations of various (+)-maltodextrin and (–)-fructose solutions confirm the Drude-Condon[8] model for Maxwell's equations in isotropic optically active media. In particular, the effective optical rotation path length, near index matching, is equal to the Goos-Hänchen shift[9] of the evanescent wave. The limits of this polarimeter, when using a continuous-wave laser locked to a stable high-finesse cavity, should match sensitivity measurements for linear birefringence ($3\times10^{-13}$ rad)[10], which is several orders of magnitude more sensitive than current chiral detection limits[7], transforming the power of chiral sensing in many fields.**

Cavity-enhanced optical methods allow the amplification of spectroscopic signals by the number of cavity passes. Achieving up to $10^5$ cavity passes with effective pathlengths of up to 100 km has allowed record sensitivities for measurements of absorption and birefringence[11,12]. These spectroscopies typically use two-mirror cavities without intracavity optics. However, chiral optical rotation (O.R.) cannot be measured in such cavities, as the O.R. from a cavity pass ($+\phi_C$) and from the returning pass ($-\phi_C$) cancel (see Fig. 1). In addition, spurious linear birefringence (e.g. in the mirrors, cell windows, or the sample), which can be considerably larger than the chiral signals, are also amplified by the cavity.

Evtuhov and Siegman[13] and Kastler[14] proposed that the cancellation of chiral O.R. in a cavity can be avoided by placing the chiral sample between two intracavity quarterwave plates, and Poirson et al. [15] performed qualititative gas-phase O.R. measurements with such a cavity. Vaccaro and coworkers demonstrated a solution to the problem of linear birefringence[16,17]: they offset the optical axes of the quarter waveplates by an angle $\alpha$, which introduces a *circular* birefringence of $2\alpha$ per round-trip pass (typically $\alpha \approx 5°$). For a laser pulse introduced into the cavity of length d, the round-trip time $t_r=(2d/c)$, where c is the speed of light. The linear polarisation rotates as the light traverses the cavity, with frequency $\omega_0 = 2\alpha/t_r = \alpha(c/d)$. This polarisation rotation suppresses the effects of spurious linear birefringence $\eta$ (as long as $2\alpha \gg \eta$), as the polarisation samples both positive and negative linear birefringence equally (on both sides of the optical axis of the birefringence). Therefore, the total acquired linear-birefringent phase shift for a complete polarisation rotation is 0. Upon adding a chiral sample, the polarisation rotation frequency becomes $\omega_C = (\alpha+\phi_C)(c/d)$. Measurement of $\omega_0$ (for an empty cavity) and $\omega_C$ (with the chiral sample) allows the determination of $\phi_C$. The main limitation is the required removal of the sample, which makes the background subtraction slow, and difficult or impossible in some cases.

To achieve the absolute measurement of chirality, we demonstrate a polarimeter based on a bowtie ring cavity[18]. Unlike a two-mirror cavity, a ring cavity can support two distinct counter-propagating laser beams (see Fig. 1), which we denote as clockwise (CW) and counter-clockwise (CCW). The symmetry between CW and CCW is broken by a longitudinal magnetic field **B** applied to an intracavity magneto-optic window, which induces a Faraday rotation $\theta_F$. A chiral sample is introduced to only one arm of the cavity. Faraday and chiral optical rotations have different symmetries: $\theta_F$ is only determined by **B**, and has the same sign for CW and CCW, whereas $\phi_C$ is only determined by the propagation direction, and has opposite signs for CW and CCW (definitions are given for the laboratory frame; see Fig. 1). Therefore, the total single-pass optical rotations for CW and CCW are given by the sum and the difference of $\theta_F$ and $\phi_C$, respectively: $\Theta_{CW}= \theta_F+\phi_C$, and $\Theta_{CCW}= \theta_F-\phi_C$. Traversing through the cavity, the light polarisation rotates with angular frequencies $\omega_{CW}(\pm B) = (\pm\theta_F+\phi_C)c/L$ and $\omega_{CCW}(\pm B) = (\pm\theta_F-\phi_C)c/L$, where the dependence of $\omega_{CW}$ and $\omega_{CCW}$ on the sign of B is shown explicitly, and L is the total round-trip cavity length. The difference $\Delta\omega(\pm B) = |\omega_{CW}(\pm B)| - |\omega_{CCW}(\pm B)|$ yields $\pm 2\phi_C(c/L)$. This key result shows that inverting the sign of B inverts the sign of the measured angle $\phi_C$ (see Fig. 2). Utilizing this signal reversal yields: $\Delta\omega(+B) - \Delta\omega(-B) = 4\phi_C(c/L)$. For each subtraction, the chiral signal $\phi_C(c/L)$, which is odd under reversal of the light propagation direction or of B, doubles. In contrast, all background signals, which are even under either reversal, cancel.

These signal reversals allow the measurement of chiral optical rotation in the presence of large backgrounds, as demonstrated below in three different environments: (i) pressure controlled and (ii) open-air chiral vapours, and (iii) chiral solutions at a prism surface probed using evanescent waves. Experiment (i) demonstrates the full symmetry of the signal reversals in the absence of large noise, experiments (ii) and (iii) take advantage of these signal reversals to measure chiral OR in high-noise environments.

For experiment (i), various pressures of (+)-α-Pinene and (−)-α-Pinene are introduced into an intracavity cell, and the four polarisation frequencies $\omega_{CW}^{+B}, \omega_{CW}^{-B}, \omega_{CCW}^{+B}$, and $\omega_{CCW}^{-B}$ are measured (Fig. 2a). The angle $\phi_C$ is determined for both (+) and (−) enantiomers, and for both +B and −B magnetic fields, and is plotted versus pressure in Fig. 2b. The optical rotation $\phi_C$ varies linearly with pressure, and the expected symmetry is obtained: $\phi_C$ reverses sign for each inversion of enantiomer or B field. Notice how an inversion of the B field allows the determination of absolute optical rotation angle, without needing to change the gas pressure.

For experiment (ii), we perform open-air measurements by inserting and removing a tray filled with liquid (+)-α-Pinene or (−)-α-Pinene, below one of the arms of the cavity, and measuring the optical rotation of the vapor that evaporates. The four frequencies $\omega_{CW}^{+B}, \omega_{CW}^{-B}, \omega_{CCW}^{+B}$, and $\omega_{CCW}^{-B}$ were measured and shown in Fig. 3a. We see that each of the four traces separately yields incorrect results for the optical rotation, some even giving the wrong sign. Strong variations in the index of refraction of the vapour alter the alignment of the cavity, yielding spurious changes in the polarisation beating frequencies, which are larger than those from the optical rotation. Also, a temperature drift causes a downward slope in all four channels. However, the result of the two signal reversals, shown in Fig. 3b, yields a constant null signal (tray removed) and measurement of optical rotation of the open-air (+)-α-Pinene and (−)-α-Pinene vapours. Comparing to Fig. 2b, we deduce that the partial pressure of the vapours was about 4 mbar, in agreement with the vapour pressure of α-Pinene.

Finally, for experiment (iii), we insert a dove prism into the beam, such that the laser beams undergo total internal reflection (TIR), with angle of incidence θ=84° (see Figs. 1 and 4). Figure 4a shows $\omega_{CW}^{+B}, \omega_{CW}^{-B}, \omega_{CCW}^{+B}$, and $\omega_{CCW}^{-B}$. Time-dependent birefringent variations in the prism cause drifts in the polarisation oscillations, which mask chiral O.R. signals in any single trace. Figure 4b shows the chiral O.R. signal obtained from the two reversals, which now show clear O.R. signal differences between the three solutions. Figure 4c shows the dependence of the maltodextrin and fructose O.R. signals on the solution refractive index n, and emphasizes a strong increase as n approaches $n_{critcal}$. An analytical expression is derived for the optical rotation from a chiral sample in an evanescent wave, according to the Drude-Condon model for Maxwell's equation in isotropic optically active media, using the treatment developed by Silverman[19,20]:

$$\phi_{EW} \simeq \frac{\Delta n}{n} \frac{N}{1-N^2} \frac{\cos\theta}{\sqrt{\sin^2\theta - N^2}} \quad (1)$$

where θ is the incidence angle, Δn = ($n_+$-$n_-$), n=($n_+$+$n_-$)/2, $n_+$ and $n_-$ are the refractive indices of the chiral sample for left and right circularly polarised light, respectively, N=(n/$n_p$), and $n_p$=1.453 is the refractive index of the prism. The data agree well with theoretical predictions, which are calculated from Eq. (1), using $\Delta n_M$ = 4.25x$10^{-6}$ $c_M$ = (2.665n-3.545)x$10^{-5}$ for maltodextrin, and $\Delta n_F$ = −2.28x$10^{-6}$ $c_F$ = (2.635-1.955n)x$10^{-5}$ for fructose (determined from single-pass O.R. measurements through a 10 cm cell), where $c_M$ and $c_F$ are the concentrations (g/$cm^3$) of the maltodextrin and fructose solutions, respectively. Notice that $\phi_{EW}$ increases sharply near critical angle (N ≈ sinθ), and even more so as N approaches 1 (near index matching, when also sinθ≈1). We approached index matching closely with N= (1.442/1.453) = 0.9924.

To better understand the physics behind Eq. (1), we express $\phi_{EW}$ in terms of the Goos-Hänchen shift $L_{GH}$: $\phi_{EW} = (\pi/\lambda)(\Delta n/[1-N^2])(\cos^2\theta/\sin\theta)L_{GH}$, where $L_{GH} = \lambda\tan\theta/\left(\pi n_p \sqrt{\sin^2\theta - N^2}\right)$ is the shift of the light beam at total internal reflection[9] (see Fig. 4d), which is the expected relevant length scale for evanescent-wave optical rotation. For an equivalent transmission measurement the optical rotation would be $\phi_T = (\pi \Delta n/\lambda) L_{eff}$, where $L_{eff}$ is the effective pathlength. Setting $\phi_{EW}$ and $\phi_T$ equal gives $L_{eff} = \{\cos^2\theta/[\sin\theta(1-N^2)]\} L_{GH}$. Away from critical angle $L_{eff} < L_{GH}$. Near critical angle and index matching (N≈sinθ≈1), $L_{eff} \approx L_{GH}$, showing that in this case the Goos-Hänchen shift is the effective evanescent-wave optical rotation pathlength.

The cavity-enhanced polarimetric methods presented here can be extended to a continuous-wave laser locked to a stable high-finesse cavity, which is much more sensitive[10] but also experimentally more complex. This should extend chiral sensitivities for conventional O.R. and circular dichroism (C.D.) measurements by several orders of magnitude, and allow the routine analysis of sub-nanoliter volumes. Applications include the study of surface chirality (for which large effects have been recently shown) [21], coupling of O.R. and C.D. to gas and liquid chromatography for sensitive chiral analysis, monitoring of protein folding, and measurement of parity violation in atoms and molecules for which insufficient pathlengths are otherwise available[22].

**Methods** An 800 nm 35 fs laser pulse is inserted into both CW and CCW directions. The mirror reflectivities R ≈ 99.7% and the cavity length L=3.6 m. All intracavity optics were antireflection coated for 800 nm (with reflectivities below 0.01%). The gas cell and the open-air tray are both of length $l_0$=0.75 m. The 70° dove prism has dimensions 80x25x25 mm. The time-dependent intensity of the output light decays exponentially[16] as $e^{-t/\tau_0}$, where the photon lifetime $\tau_0$=L/[c(1-$R^4$)] ≈1 μs (see Fig. 2a). Faraday rotation $\theta_F \approx$ 2.5-4° was generated by applying a 0.2-0.3 T magnetic field to a 3 mm thick terbium gallium garnet (TGG) crystal. Using a polariser at the output, the optical rotation appears in the CW signal $I_{CW}$ as an oscillation with frequency $\omega_{cw}$: $I_{cw} = I_0 e^{-t/\tau_0} \cos^2[\omega_{cw} t/2]$, and in the CCW signal with frequency $\omega_{CCW}$: $I_{ccw} = I_0 e^{-t/\tau_0} \cos^2[\omega_{ccw} t/2]$, where $I_0$ is the output intensity at t=0 (see Fig. 2a). Each data point represents five sets of 4000 laser shots. The data traces are fit with the $I_{CW}$ and $I_{CCW}$ fit functions, to determine $\omega_{cw}$ and $\omega_{ccw}$. The magnetic field was reversed between each data point.

Enantiopure (+)-α-Pinene or (−)-α-Pinene (Sigma Aldrich) were used. Maltodextrin and fructose were bought commercially. Maltodextrin and fructose solutions (with concentrations between 50-60%) and glycerol/water solutions are prepared with refractive indices *n* ranging from 1.417 to 1.442 (at 0.005 intervals,). For the evanescent-wave setup (Fig. 1b iii) a magnesium fluoride compensator was inserted to reduce the birefringence δ of the prism (typically 10-20°) to about 0.5° for both beams (the large position-sensitive birefringence in the prism and imperfect beam alignment precluded better compensation). Modeling the depolarisation effects of birefringence[22], the ratio $δ/θ_F < 0.2$ yields a correction coefficient $q^2 > 0.99$, so that the effects of birefringence are small.

**Acknowledgements** This research was supported by the ERC grant TRICEPS (grant no. 207542), and the FP7 IAPP Programme SOFORT (PIAPGA-2009-251598). We thank Dr. Paris Tzallas for access to the Attosecond labs at IESL-FORTH.

**Author Contributions** L.B. constructed the experiment, performed the gas-cell and open-air experiments, and analysed the data. D.S. performed the evanescent-wave experiments, and analysed the data. G.E.K. and A.K.S. developed the data acquisition and analysis software, and assisted in the experiments. G.E.K. prepared the figures. B.L. derived the evanescent-wave optical rotation equations. T.P.R. conceived of and directed the experiments, and wrote the manuscript. All authors provided important suggestions for the experiments, discussed the results and contributed to the manuscript.

**Author Information** Reprints and permissions information is available at www.nature.com/reprints. The authors declare no competing financial interests. Readers are welcome to comment on the online version of the paper. Correspondence and requests for materials should be addressed to T.P.R. (ptr@iesl.forth.gr).

**Figure 1 | Cavity-enhanced polarimeter for chiral sensing**. **a**, The layout of four mirrors (M1-4), polarisers (P1, P2), photomultiplier tubes (PMT1, PMT2), the chiral sample, Faraday medium, and the counter-propagating CW and CCW beams. **b**, The chiral sample gives opposite laboratory-frame optical rotations for CW and CCW beams ($\phi_C^{CW} = -\phi_C^{CCW}$). Measurements were performed on chiral samples in a gas cell, open air, and in an evanescent wave at a prism surface. **c**, The Faraday rotator gives the same optical rotations for CW and CCW beams ($\theta_F^{CW} = \theta_F^{CCW}$), controlled by the magnitude and sign of the applied magnetic field **B**.

**Figure 2 | Gas-cell optical rotation**. **a**, Experimental signals showing the polarisation oscillation frequencies $\omega_{CW}^{+B}, \omega_{CW}^{-B}, \omega_{CCW}^{+B}$, and $\omega_{CCW}^{-B}$ in the exponential decay, for 4 mbar and 0 mbar of (–)-α-Pinene vapour. Notice the sign change of the frequency shift between CW and CCW or +B and –B signal pairs (insets). **b**, Measurements of optical rotations $\phi_C$ for +B and –B, and for (+)-α-Pinene and (–)-α-Pinene vapours, as a function of gas pressure. Error bars are 2σ confidence intervals.

**Figure 3 | Open-air optical rotation.** **a**, The four polarisation frequencies $\omega_{CW}^{+B}, \omega_{CW}^{-B}, \omega_{CCW}^{+B}$, and $\omega_{CCW}^{-B}$, are shown for open-air measurements of (–)-α-Pinene vapour, evaporating from a tray which is periodically inserted and removed. Each of the four polarisation frequencies is dominated by spurious signals and background drifts. **b**, Subtractions of the polarisation signals yield the open-air optical rotation, shown for (+)-α-Pinene and (–)-α-Pinene vapours, and a stable background. The 2σ statistical error bars are determined from replicate measurements.

**Figure 4 | Evanescent-wave optical rotation.** **a**, The four polarisation frequencies $\omega_{CW}^{+B}, \omega_{CW}^{-B}, \omega_{CCW}^{+B}$, and $\omega_{CCW}^{-B}$, are shown for evanescent-wave measurements of maltodextrin, glycerol, and fructose solutions, all with refractive index n=1.442. The solutions are changed with a flow cell every 10 minutes. **b**, Subtractions of the polarisation signals give the optical rotation of the samples in the evanescent wave. **c**, Measurements for solutions with n=1.417-1.442. Error bars are 2σ confidence intervals. The theoretical curves are generated using Eq. (1). **d**, The Goos-Hänchen shift $L_{GH}$ of the evanescent wave.

Figure 1:

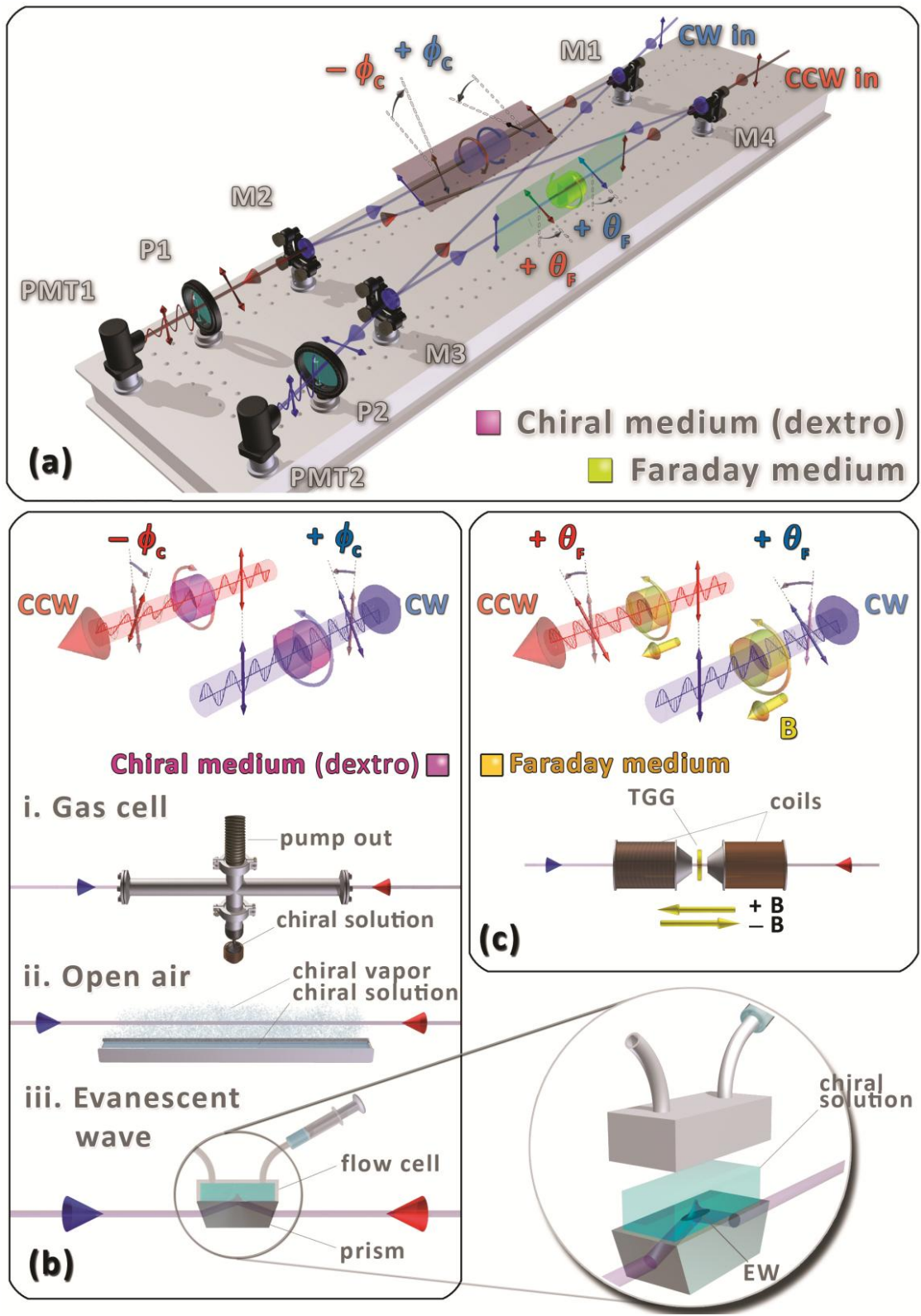

Figure 2:

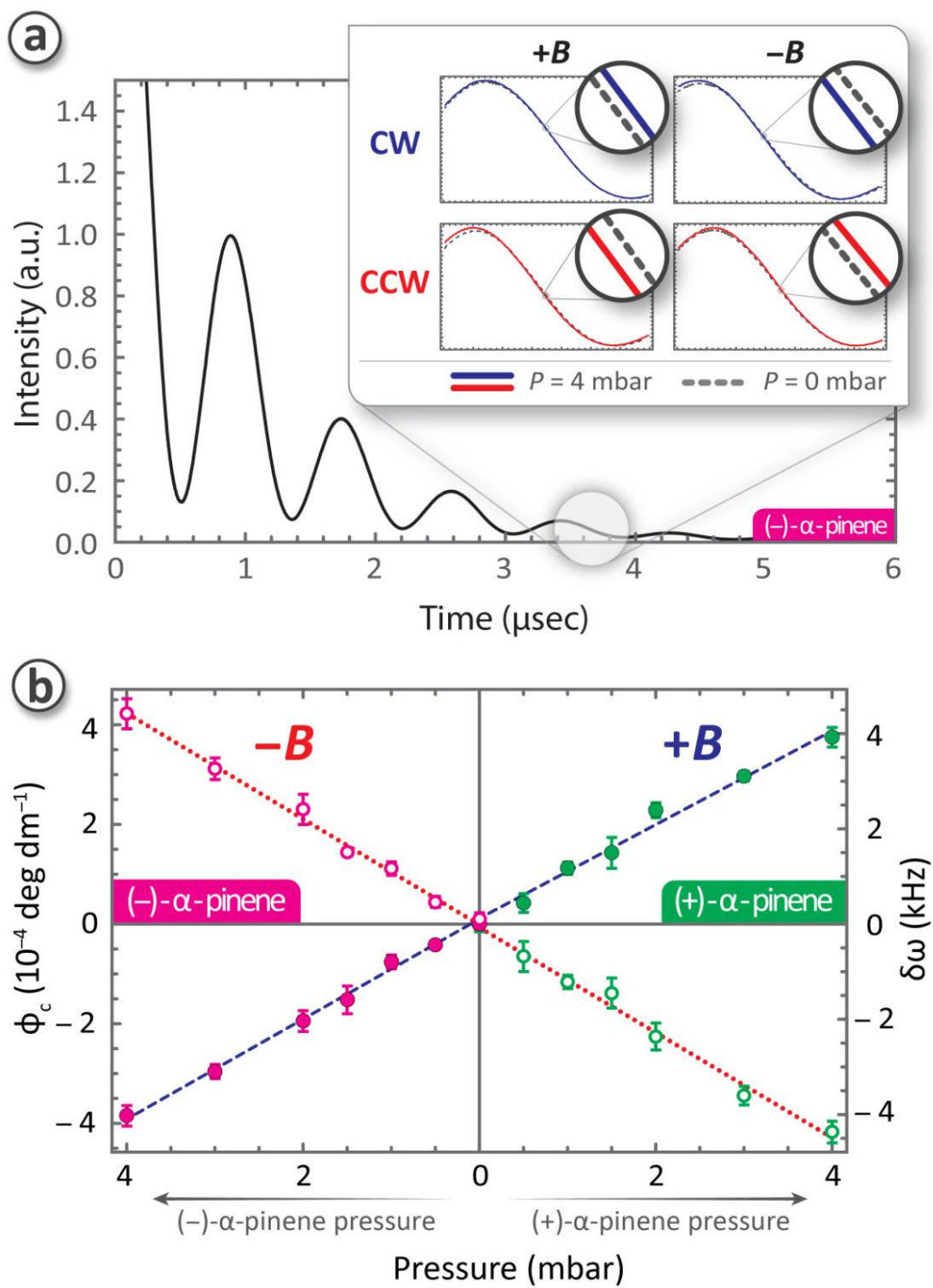

Figure 3:

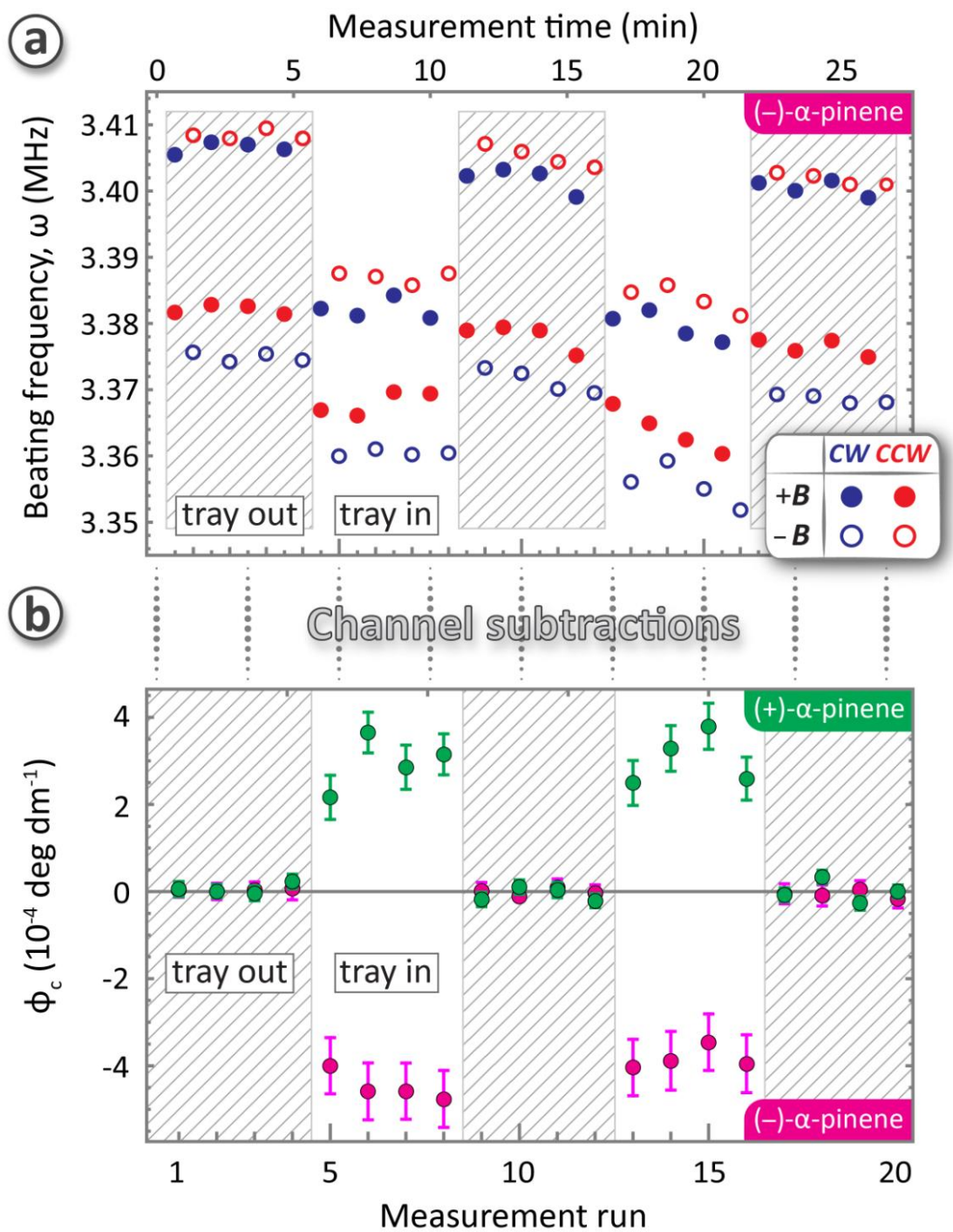

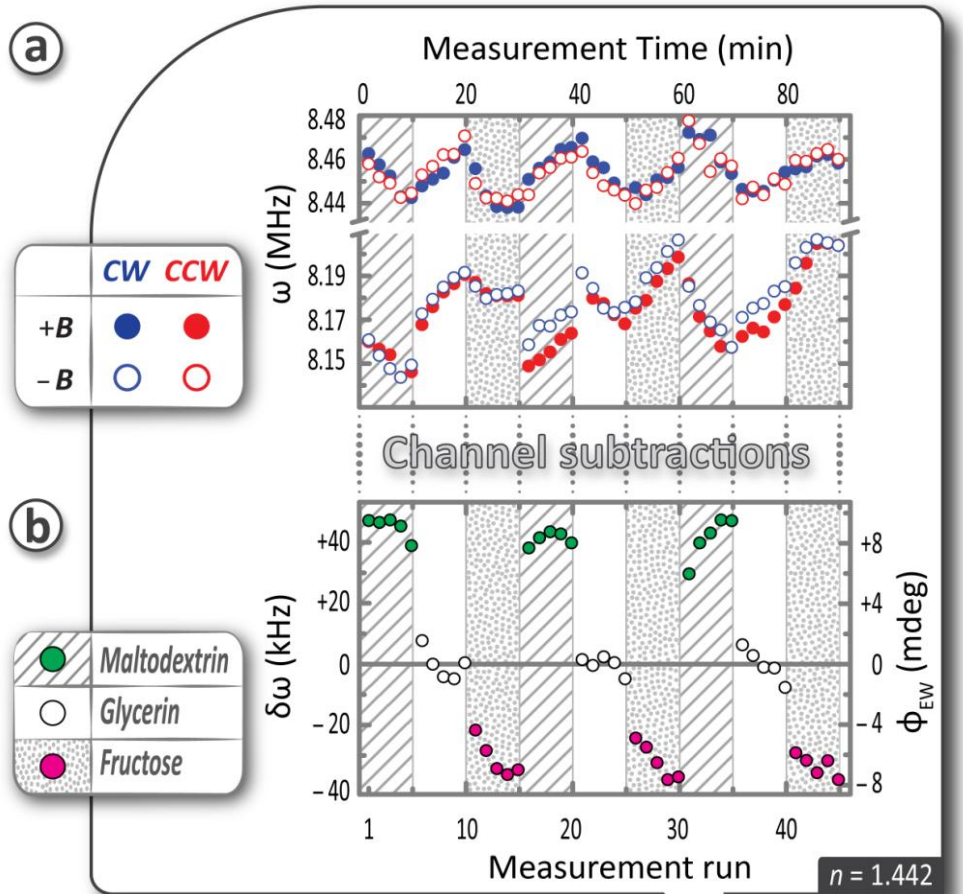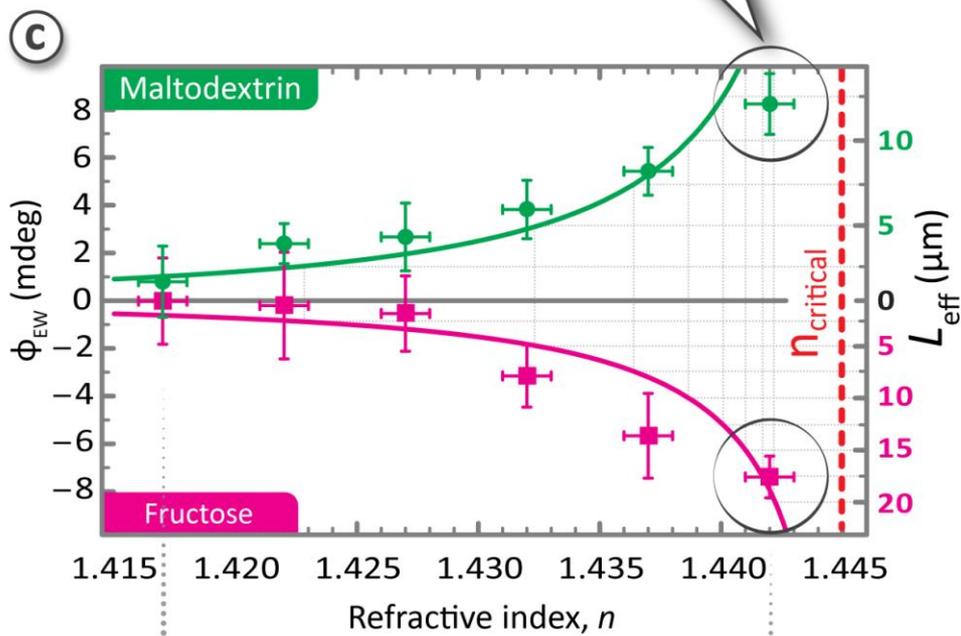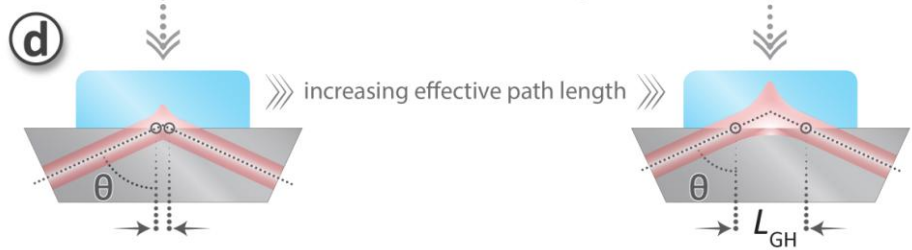